# INVESTIGATION OF FAINT GALACTIC CARBON STARS FROM THE FIIRST BYURAKAN SPECTRAL SKY SURVEY.   III. INFRA – RED CHARACTERISTICS

K. S. Gigoyan[1], A. Sarkissian[2], C. Rossi[3], D. Russeil[4], G. Kostandyan[1],

M. Calabresi[5], F. Zamkotsian[4],  M. Meftah[2]

MI.

[1] V. A. Ambartsumian Byurakan Astrophysical Observatory, Armenia,     e-mail: kgigoyan@bao.sci.am

[2] Universite de Versailles Saint-Quentin, CNRS/INSU, LATMOS-IPSL, France,

   e-mail: Alain.Sarkissian@latmos.ipsl.fr , Mustapha.Meftah@latmos.ipsl.fr

[3] Dipartimento di Fisica, University di Roma La Sapienza, P. le Aldo Moro 5, I-00185 Roma Italy,

   e-mail: Corinne.Rossi@roma1.infn.it

[4] Laboratoire d Astrophysique de Marseille, CNRS-AMU, France,

   e-mail: delphine.russeil@lam.fr, frederic.zamkotsian@lam.fr

[5] Associazione Romana Astrofili - Frasso Sabino Observatory, Italy,        e-mail : m.calabresi@mclink.it

**Abstract**

  Infra-Red(IR) astronomical databases, namely, IRAS, 2MASS, WISE, and Spitzer, are used to analyze photometric data of 126 carbon(C) stars whose spectra are visible in the First Byurakan Survey(FBS) low-resolution(lr) spectral plates. Among these, six new objects, recently confirmed on the digitized FBS plates, are included. For three of them, moderate-resolution CCD optical spectra are also presented. In this work several IR color-color diagrams are studied. Early and late-type C stars are separated in the JHK Near-Infra-Red(NIR) color-color plots, as well as in the WISE W3-W4 versus W1-W2 diagram. Late N-type Asymptotic Giant Branch(AGB) stars are redder in      W1-W2, while early-types(CH and R giants) are redder in W3-W4 as expected. Objects with W2-W3 > 1.0 mag. show double-peaked spectral energy distribution(SED), indicating the existence of the circumstellar envelopes around them. 26 N-type stars have IRAS Point Source Catalog(PSC)     associations. For FBS 1812+455 IRAS Low-Resolution Spectra(LRS) in the wavelength range 7.7÷22.6μm and Spitzer Space Telescope Spectra in the range 5÷38μm are presented clearly showing absorption features of $C_2H_2$(acetylene) molecule at 7.5 and 13.7μm, and the SiC(silicone carbide) emission at 11.3μm. The mass-loss rates for eight Mira-type variables are derived from the     K-[12] color and from the pulsation periods. The reddest object among the targets is N-type C star     FBS 2213+421, which belong to the group of the cold post-AGB R Coronae Borealis(R CrB) variables.

*Key words: stars: carbon stars: Mira-type stars: infra-red characteristics: mass-loss rate*

1. *Introduction*.

This is the third paper of the series[1, 2] devoted to the study of carbon(C) stars discovered on the First Byurakan Survey(FBS)[3] low-resolution(lr) spectral plates at high Galactic latitudes[4-6], more than 10° above the Galactic plane. Since 2007, all the FBS lr spectral plates are digitized, and the Digitized First Byurakan Survey(DFBS)[7] database was created(available online at http://ia2.oats.inaf.it, and at http://www.aras.am/Dfbs/dfbs/html for FBS zones statistics and technical data). 120 new C stars and numerous M-type stars of late-subclasses have been discovered[8]. The first two papers of this series were devoted to study the optical variability, K-band absolute magnitudes, and distance estimations of 54 FBS N-type AGB C stars, 66 C stars, showing early-type features. The detection ranges were also estimated in FBS for each group of the C-rich objects: N-type stars, CH and R stars, and dwarf carbon(dC) stars. The C-rich nature for all 120 detected stars has been confirmed by moderate-resolution CCD spectroscopy[4-6, 8]. The magnitudes of FBS C stars are in the range 12.0÷16.0 mag. in V-band. These objects are between moderately faint(fainter than 11.0÷12.0 mag.) C stars and extremely faint and distant C stars, found in the Sloan Digital Sky Survey(SDSS) commissioning data[9,10]. In this paper we present data for 6 new early-type C stars found in the DFBS database. These objects should be added to the 120 FBS+DFBS C stars already known[4-6, 8]. he goal of this paper is to analyze all possible IR data from modern astronomical catalogs for all 126 C stars. For all confirmed stars we have considered IR color-color diagrams. IRAS Low-Resolution Spectra(LRS)[11] in the wavelength range 7.7÷22.6μm and Spitzer Space Telescope Spectra in the range 5÷38μm are presented for FBS 1812+455. Relations between K-[12] color index and pulsation periods(P) are used to estimate the mass loss rate for eight C Mira-type variables. All DFBS lr spectral plates(2000 Kodak IIAF, IIF, IIaF, and 103aF plates, 4°×4° each in size) have been analyzed twice with help of standard image analysis softwares(FITSView and SAO Image ds9). This visualization allows to detect very red candidate stars close to the detection limit in each DFBS plate[6]. This visualization yielded to the discovery of numerous late-subclasses of M stars, and 6 new C stars, showing early-type characteristics. The selection criteria of late-type stars on DFBS plates are described with more details in papers[4,8]. Particularly, C stars can be identified through the presence of the Swan bands of the $C_2$ molecule at λ 4737, 5165 and 5636A(N-stars). Early-type C stars show $C_2$ absorption band at λ 4382Å also. Note that, based on recent detection of the new faint late-type stars at high latitudes, we are planning to present the 2nd Version of the "Revised And Updated Catalogue of The First Byurakan Survey of Late-Type Stars"[8].

*2. New Confirmed DFBS C Stars.*

*2.1. 2MASS Data.* Table 1 present DFBS number, 2MASS(Two Micron All-Sky Survey[13](online available at http://irsa.ipac.caltech.edu/Missions/2mass.html/) JHKs photometric data for the 6 new early-type C stars. For a possible proper motion, they were checked in optical multi-color and multi-epoch databases, like the PPMX(Catalog of Position and Proper Motions of the ICRS)[14], online access at http://vo.uni-hd.de/ppmxl) and in SuperCOSMOS Sky Survey-SSS(online at http://www-wfau.roe.ac.uk/sss/). No detectable proper motion was found for the confirmed C stars. As we can see in Table 1, NIR colors for the 6 stars are typical for early-type C giants, properties to be validated in next section.

TABLE 1    2MASS JHKs Data For 6 New Early-Type DFBS C Stars

| DFBS Number | 2MASS Association | J(mag) | J-H(mag) | H-Ks(mag) | B°(Gal. lat.) |
|---|---|---|---|---|---|
| J030610.42+435320.8 | J030610.42+435320.8 | 10.513 | 0.805 | 0.322 | - 12.568° |
| J032659.76+385650.0 | J032659.77+385649.5 | 9.415 | 0.769 | 0.276 | - 14.603 |
| J151825.96+130424.6 | J151825.84+130423.5 | 11.376 | 0.389 | 0.158 | + 52.918 |
| J163117.35+152902.3 | J163117.33+152902.2 | 11.732 | 0.814 | 0.269 | + 37.889 |
| J175212.89+341126.5 | J175212.86+341126.2 | 10.437 | 0.701 | 0.191 | + 26.382 |
| J214733.89+154104.1 | J214733.89+154104.1 | 11.932 | 0.602 | 0.133 | - 28.182 |

***2.2 Optical Spectroscopy***. For the stars DFBS J030610.42+435320.8 and DFBS J032659.76+ 385650.0 medium-resolution CCD spectra in the range 3900-8500Å (dispersion is 3.9A/pix) were obtained on 12/13 January 2016, with the 1.52m Cassini telescope of the Bologna Astronomical Observatory at Loiano(Italy, equipped with the Bologna Faint Object Spectrometer and Camera-BFOSC, 1300×1340 pix EEV P129915 CCD). Spectroscopic data were reduced by means of standard IRAF[1] procedure. Moderate-resolution CCD spectra for DFBS J151825.96+130424.6 were obtained on 15 April 2016 at the 2.6 m telescope of the Byurakan Astrophysical Observatory(BAO, Armenia, equipped with the SCORPIO spectrograph and TK 1024×1024 pixel CCD, dispersion 3A/pix.). The 1.52 m Loiano and the 2.6 m BAO telescope spectr shown in Fig. 1, are typical CH giants[15] confirming 2MASS NIR color properties. The C-rich nature for the remaining 3 objects of Table 1(for which IBI > 20°) was confirmed on Hamburg Quasar Survey(HQS) Low-Resolution Digitized database(http://www.hs.unihamburg.de/DE/For/Exg/Sur/hqs/online/index.html, HQS spectral resolution is better than DFBS's one). Fig. 1 illustrates also HQS Digitized 2D spectra for the remaining three objects of Table 1, where the absorption bands of $C_2$ molecule are very well expressed.

*3. IR-Data.*

This paper is mainly dedicated to the characterization of our targets with the help of  R data by applying the color-magnitude and color-color diagrams to data downloaded from several ground based and satellite surveys. We used the JHKs photometry from 2MASS catalog, the IRAS photometric data, which were used to select the original BIS(Vizier catalogue III/237A) sample, the observations of the AKARI satellite at 9 and 18μm[16] and of the WISE satellite at 3.4, 4.6, 12 and  22 μm(Vizier Catalogue II/328). The WISE 4 bands photometry provide useful color indices: we show in Fig. 2(a) W1-W2 versus W3-W4 and in Fig. 2(b) W1-W2 versus W1-W4. Dusty C stars are well separated in a rising branch, while non-dusty C stars are mixed with the other ones. Mira vari- ables are a bit above the main locus of the late type stars. Semi-regular variables are spread all  along the main locus, while non-variable stars are grouped in the blue corner. Similar useful tools are the plots J-Ks versus Ks-AKARI9, built using 2MASS and AKARI Catalogs(see Fig 3a), and  J-Ks versus Ks-W4(see Fig. 3b) built using 2MASS and WISE data: in both plots the C dusty stars are well separated from the bulk of the naked late type stars. We note that only the brightest dusty carbon stars from FBS are present in the AKARI database.

---

[1] *IRAF is distributed by the NOAO which is operated by AURA under contract with NFS.*

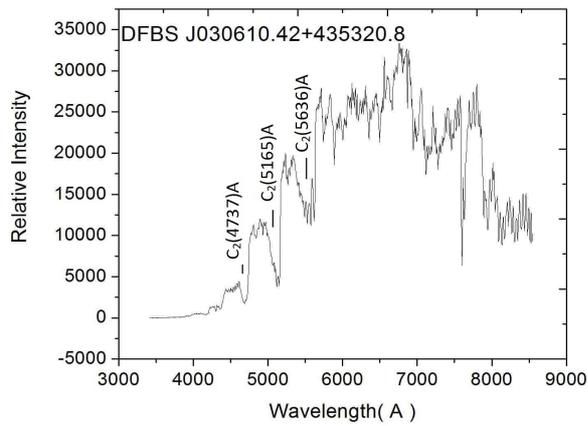
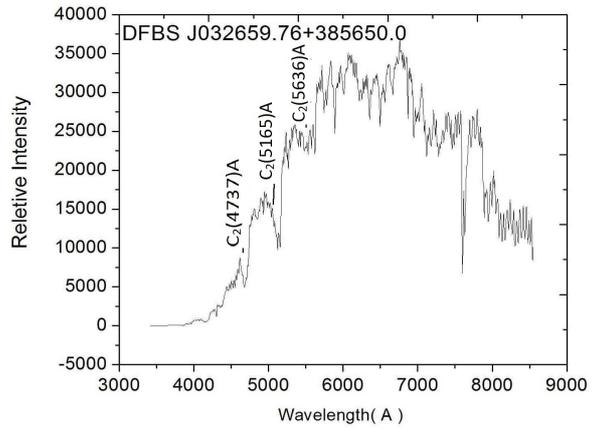
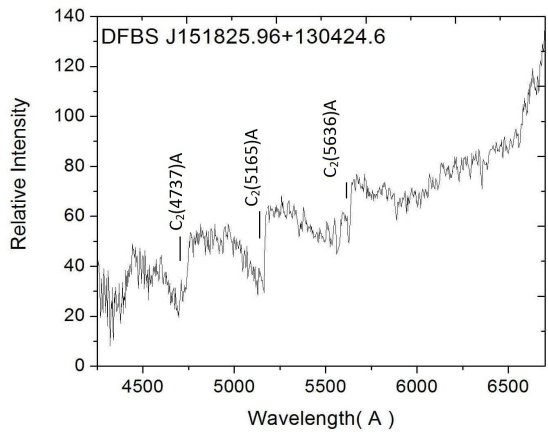
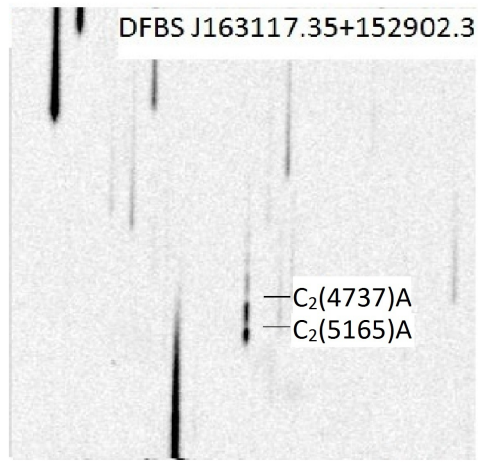
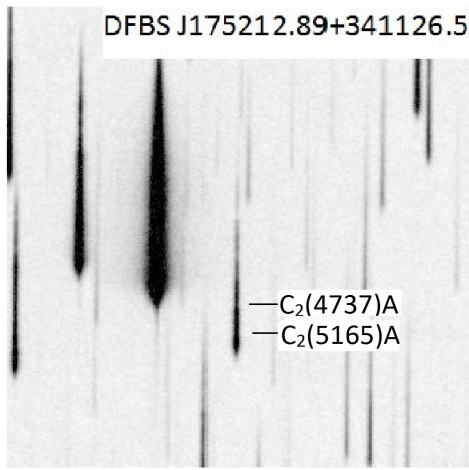
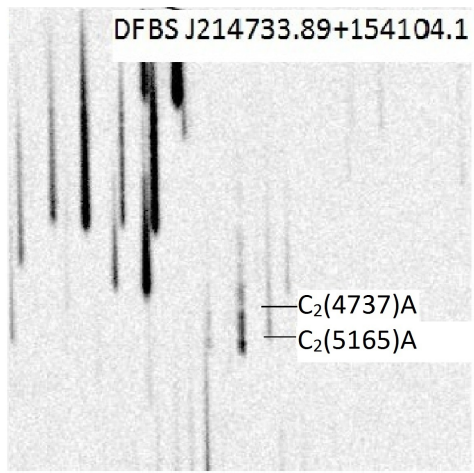

**Figure 1.** Loiano 1.52 m telescope spectra for DFBS J030610.42+435320.8 and DFBS J032659.76+385650.0 in the wavelength range 3900-8500A and 2.6 m BAO telescope spectra for star DFBS J151825.96+130424.6 in range 4250-6750A,. HQS 2D lr spectral images(each size 5°×5°) for 3 objects from Table 1. The absorption bands of $C_2$ molecule are indicated.

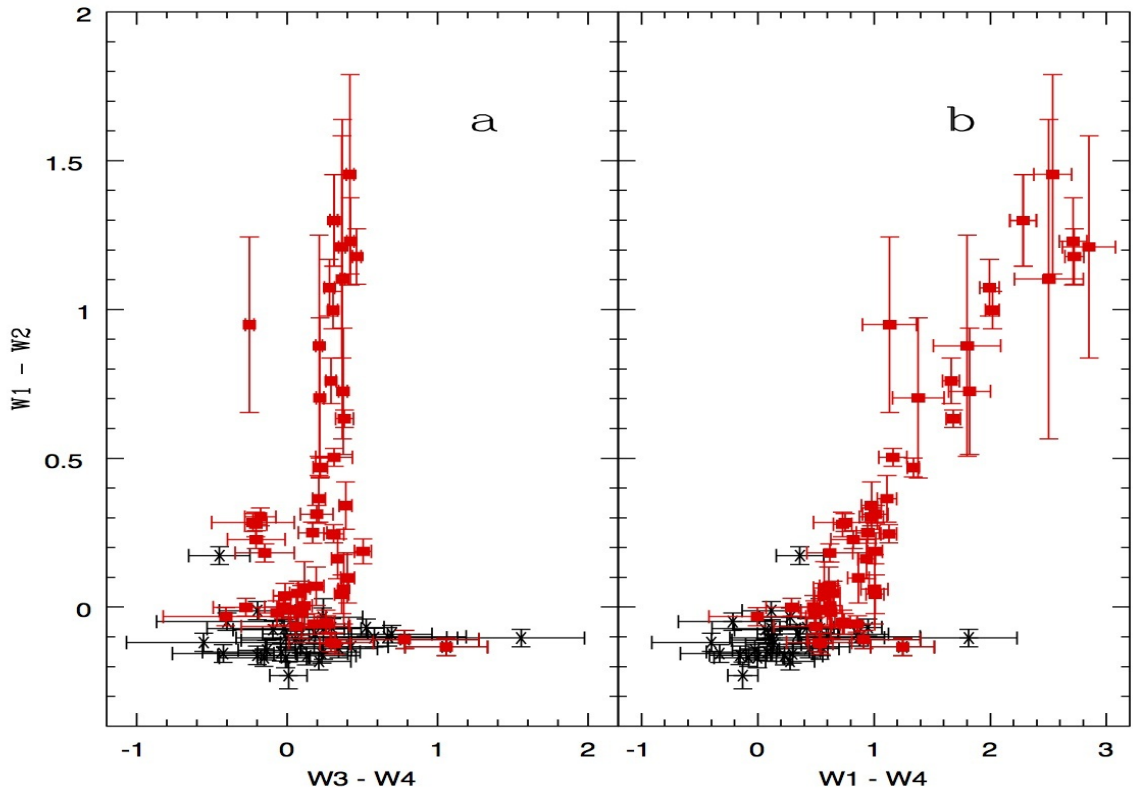

**Fig.2** WISE W1-W2 vs. W3-W4(Fig. 2a) and WISE W1-W2 vs. W1-W4(Fig. 2b) color-color plots(with error bars) for all C stars. Crosses are early-type stars, filled squares, are late N-type AGB stars.

**3.1. IRAS Data**. After IRAS mission, the IRAS data became an important tool in the infrared for studying late stages of stellar evolution. IRAS two-color plots are used in many papers and the Low-Resolution Spectra are important in discriminating between oxygen-rich(M stars) and carbon- rich(C stars). 26 objects (N-type stars only[8]) out of 126 FBS C stars detected are associated with IRAS sources[17]. Only four objects have fluxes up to 60 micron reported with good quality factors in the IRAS catalogue FBS 0137+400, FBS 0707+270, FBS 1812+455, and FBS 2213+421. We have checked the position of these stars in the classical diagram by van der Veen & Habing[18] where the colors are defined as follows:

$$[12]-[25] = 2.5 \log F(25)/F(12) \qquad (1)$$

$$[25]-[60] = 2.5 \log F(60)/F(25) \qquad (2)$$

with F(12), F(25), and F(60) being the IRAS fluxes at 12, 25, and 60μm, respectively. After color correction for the temperature to the fluxes published in the catalog, the results are those reported in Table 2. In the scheme of Figure 5b by van der Veen and Habing[18] FBS 0137+400 lie close to the regi- on VII corresponding to variable carbon stars, the others, well inside that region, are long period, irregular variables as we have verified from the check of the variability types using the Catalina Sky Survey(CSS-http://nesssi.cacr.caltech.edu/DataRelease/)[19] database.

TABLE 2   IRAS Colors For 4 N-Type Carbon Stars.

| I R A S color | FBS 0137+400 | FBS 0707+270 | FBS 1812+455 | FBS 2213+421 |
|---|---|---|---|---|
| [12]-[25] | -1.41 | -1.36 | -1.32 | -0.74 |
| [25]-[60] | -1.59 | -1.64 | -1.77 | -1.55 |

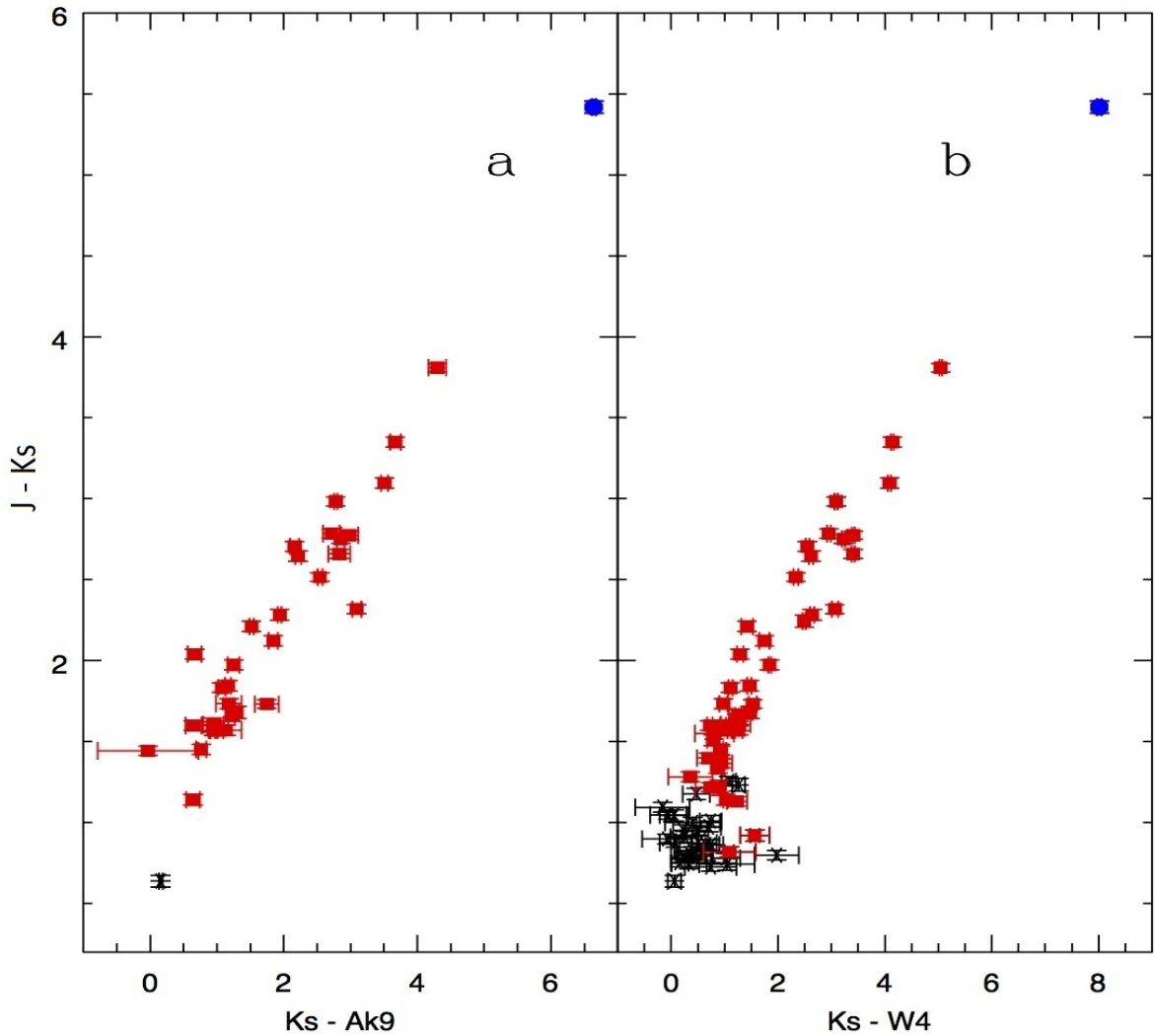

**Fig.3** 2MASS J-Ks vs. Ks-AKARI 9mag.(Fig. 3a) and J-Ks vs. Ks-W4(Fig. 3b) color- color diagram for all 126 C stars confirmed. A single(reddest) object on the upper right corner present the position of the star FBS 2213+421(filled polygon). Symbols are the same, as in Fig. 2 (a,b).

From the spectroscopic point of view, the silicate features at 10 and 18μm in emission or in absorption are indicators for belonging of objects to O-rich group. Spectral types of the large amount of unassociated IRAS Point Sources are presented also in huge amount of papers. Particularly, data for numerous of the new Infrared Carbon Stars(ICS)[20] are identified on the basis of the presence of the SiC emission feature at 11.3μm in their LRS. Only two stars(FBS 1812+455 and FBS 2213+421) have been also detected in spectroscopic mode. IRAS spectra for FBS 0137+400 and FBS 2213+421 are presented in papers[21, 22]. The long period variable FBS 1812+455 will be discussed below.

**3.2. 2MASS Colors.** To discriminate dwarf/giant luminosity class, we used the traditional color-color plots (J-H vs. H-Ks)[23]. In Fig. 3 of paper[1], we present 2MASS color-color plots for 120 FBS+DFBS C stars. This diagram clearly shows the sharp division between N giants and other C stars(the different location of early-type and late-type C stars, as in paper[2]). The reddest object, at the uppermost right corner[1] is the N-type star FBS 2213+421, which belongs to the group of the cold post-AGB R Coronae Borealis Variables(R CrB)[24].

**3.3. *Spitzer Data.*** We have checked all 126 confirmed C stars for possible detection in Spitzer database(http://sha.ipac.caltech.edu/applications/Spitzer/SHA/). Only for the late N-type star FBS 1812+455, which is a Mira-type variable, the Spitzer Infrared-Spectrograph(IRS) spectra[12] are available in the range 5÷38μm. We present these spectra, together with the IRAS LRS in Fig. 4(a, b). The overall spectrum of FBS 1812+455 is very similar to that of Sgr 22(2MASS J19103987-3228373=IRAS 19074-3233, a foreground member of the Galaxy)[25] which is a variable with period 370 days and has also the same 2MASS J-H and H-Ks colors. In Fig. 4b the most interesting features are the absorption bands of $C_2H_2$ at 7.5 and 13.7μm and the very strong SiC emission at 11.3 μm. This last feature is strong in Spitzer but it is also clearly detectable in the IRAS spectra.

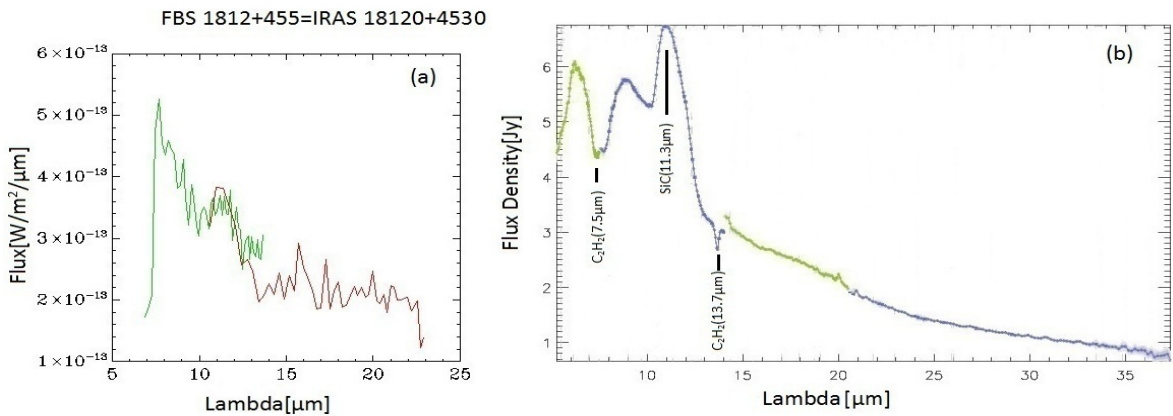

**Fig. 4** IRAS Low-Resolution Spectra (LRS) in wavelength range 7.7÷22.6 μm(Fig. 4a) and Spitzer/IRS spectra(Fig. 4b) in wavelength range 5÷38 μm for FBS1812+455.

**3.4. *WISE Data.*** WISE(Wide-Field Infrared Survey Explorer)[26] mapped the whole sky in 4 in- frared bands, W1, W2, W3, and W4 centered at 3.4, 4.6, 12, and 22μm (see CDS catalogue II/328 for photometric data, image data access is available at http://irsa.ipac.caltech.edu/applications/wise). The WISE 4 bands photometry provides useful color indexes although with some caveat. Actually, while in most of the color-color diagrams the N giants are distributed along a branch well separated from the earlier types, CH, R and dwarf carbons have the same colors as the early M type stars. The carbon nature of these stars should be also confirmed by other evidences[27]. Color-magnitude crieria based on WISE data are developed in paper[28] and allowed to select AGB stars with circum- stellar dust shells, and separate C-rich from O-rich classes. This database is analyzed also for C and M-type AGB stars in the Galaxy[29]. All the 126 FBS+DFBS C stars were detected by WISE (see Fig. 2(a,b) for WISE color-color plots).

*3.5. AKARI Data*. 39 objects(out of 126) were associated with the AKARI Point Source Catalog sources(CDS Catalogue II/298). They are 38 N-type stars plus the CH star FBS 0018+213=AKARI 0021334+213526, which is the brightest early-type CH star among the FBS sample. The star FBS 0043+474= AKARI 0046284+474132 has no record on 9μm band, but it is observed at 18μm band (S18=1.01E+00 Jy). The remaining 16 N stars were not detected by AKARI satellite, because of their faintness. The calibrated flux densities reported in the catalogue were converted into magnitudes of the IRC-Vega system using the zero-magnitude flux densities from Tanabe et al.[30]. The uncertainties were derived from the maximum and minimum magnitudes derived from the flux uncertainties. In most cases Δmag is below 0.05, for the faintest 15 cases Δmag ranges between 0.1 ÷0.15 Fig. 3(a,b) presents 2MASS J-Ks vs. Ks-AKARI 9 mag. and 2MASS J-Ks vs. Ks-W4 color-color plots. Our purpose was not only to study the distribution of the points, but also to make a comparison between the two diagrams. As we can see the agreements is very good and uncertainties are comparable. Again, the variable AGB N-type stars are distributed along the strip of progressive reddening; the single CH star in the diagram is located in the lowermost corner, as expected(Fig. 3a). Objects with color index $K_S$-AKARI9 mag. > 3.0 mag., show double-peaked spectral energy distribution(SED).

## *4. Spectral Energy Distribution-SED*.

SED for numerous of evolved late-type stars are presented in large amount of papers in our own Galaxy and in members of the local group, particularly, in Large and Small Magellanic Clouds[31]. We have constructed the SED of all our stars to check their variability in the optical and study the emission excess in the infrared. We have collected the photometry from Vizier(http://vizier.u-strasbg.fr/viz-bin/VizieR/) in the wavelength range from the B-band(0.444 μm) to the IRAS 100μm band is covered by the SED. We used 2MASS, WISE, AKARI and IRAS for the IR, UCAC4 and APASS catalogues(CDS Catalogues I/322A and II/336), for the optical bands, then applied the conversion λ*F(λ) using the tool of the web site http://morpheous.phys.lsu.edu/magnitude.html. For some stars the interstellar extinction is not negligible in various photometric bands(for example for some objects Av > 0.45 mag.). In these cases we have applied the corrections as described in paper[2]. It is important to note that the photometric data downloaded from the various catalogues are obtained at different epochs(not simultaneously). This could represent a problem for N-type AGB which are variable mainly in the optical. Actually, we have kept all the points to have an idea of the variability. Early-type type CH and R giants are more or less stable. 5 FBS N-type AGB C stars show double-peaked SED. In Fig. 5 we present the observed SED for some interesting objects.

## *5. IR-Colors And Mass-Loss Rates*.

IR colors can be used to obtain mass loss rates for N-type stars. Our estimation for eight Mira-type variables is based on IR color and pulsation period. Whitelock et. al.[32] determined the correlation between K-[12] index and mass-loss rates for Galactic C Miras. This relation has quantified as:

$$\mathrm{Log}(\dot{M}_{total}) = -7.668 + 0.7305(K-[12]) - 5.398 \times 10^{-2}(K-[12])^2 + 1.343 \times 10^{-3}(K-[12])^3 \quad (3)$$

where [12] is the IRAS 12 μm band magnitude defined as:

$$[12] = 3.63 - 2.5 \times \mathrm{Log}\, F(12) \quad (4)$$

where F(12) is the IRAS flux in Jansky at 12 μm. As a supplementary evaluation we have also used the correlation between mass-loss rate and pulsation period[33]. This correlation is defined as follow:

$$\mathrm{Log}\dot{M} = (4.08 \pm 0.41) \times \mathrm{Log}P - (16.54 \pm 1.1) \quad s=0.27 \quad (5)$$

Table 3 presents the mass-loss rates for eight Mira variables, where columns present: FBS or DFBS number, 2MASS association, and estimation of LogM, using K-[12] color indices and pulsation periods(P). Variability types and periods are given in paper[1]. As we can see(Table 3), similar values are obtained for mass-loss rates, using the two different methods. The larger difference in the values for FBS 1812+455, most probably, can be explained by K-band variability. The distances to these objects, which are in the range 3.6 to 12.9 kpc, determined and presented in paper[1], are based on Period-Luminosity(PL) relations.

TABLE 3 Mass-Loss Rates For 8 N-Type Mira Variables.

| FBS+DFBS Number | 2MASS Association | LogM(K-[12] | LogM(P) |
|---|---|---|---|
| FBS 0043+474 | 00462480+4741330 | -6.18 | -6.19 |
| FBS 0155+384 | 01580610+3839185 | -7.00 | -6.43 |
| FBS 0158+095 | 02005614+0945356 | -5.36 | -5.38 |
| FBS 0502+088 | 05050029+0856078 | -6.00 | -5.80 |
| FBS 0729+269[1] | 07323273+2647156 | -6.60 | -6.60 |
| FBS 1812+455 | 18132945+4531175 | -5.13 | -5.92 |
| DFBS J064958.64+741610.1 | 06495846+7416107 | -5.80 | -6.08 |
| DFBS J230835.19+403533.9 | 23082356+4035287 | -6.06 | -6.10 |

[1] **For FBS 0729+269 as a [12] μm magnitude the WISE W3 band magnitude is adopted.**

## 6. Discussion and conclusion.

All available IR data from modern astronomical databases are exp-loited to study C stars found on FBS spectral plates. IR observations enabled us to study dust and gas features from the circumstellar envelopes around these stars. The well-known relation between K-[12] colour and pulsation periods was used to estimate the mass loss rate for eight Mira-type variables. The values obtained are between $10^{-5}$ and $10^{-7}$ $M_{solar}$/yr. and are typical for N–type AGB variable stars. Optical and IR photometric data are used to construct the Spectral Energy Distributon for all detected C stars. 5 N-type stars with W2-W3>1.0 mag. show double-peaked SED, indicating the existence of an envelope around them. For FBS 1812+455 it is obvious from the SED, with a emission peaking around 5 micron(showing maximum radiation around 5 μ wavelength) and from Spitzer spectra. Also differences in two values of mass-loss rates for this objects is explained with large amplitude variability in K band and in IR bands. This object needs to be monitored in NIR bands and studied in more detail in future. The reddest object among the studied targets is N-type star FBS 2213+421(Fig. 3(a,b)), which belongs to the cold post-AGB R Coronae Borealis(R CrB) variables. We note, that similar work is in progress for all detected FBS+DFBS M-type stars, which are more than 1500 in the 2nd version of the of "Revised And Updated Catalogue of The First Byurakan Survey of Late-Type Stars"[8]. Part of our program also is to obtain a high spatial resolution NIR and IR images for comparatively bright and interesting stars.

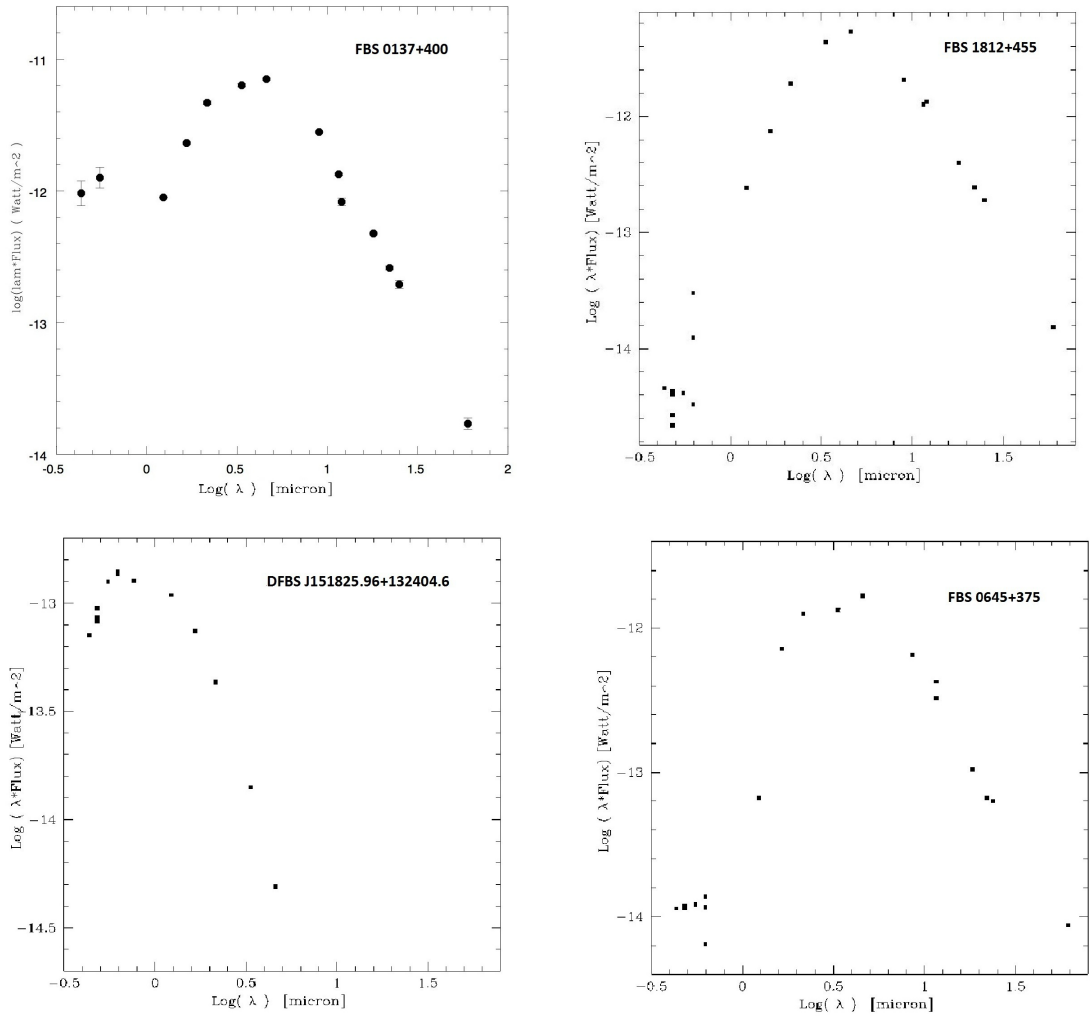

**Fig. 5.** Spectral Energy Distribution (SED) for some stars. X-axis presents wavelength in micron, Y-axis presents flux in units Watt/m² .

K.S.G. thanks CNRS, LATMOS, University of Versailles Saint Quentin en Yvelines, and LAM for supporting this study. This works was supported also by the RA MES State Committee of Science, in the frames of the research project No. 14PR-1C0109. Authors thank the staff of the Cassini telescope for technical assistance during the observations. This research has made use of the SIMBAD and VIZIER data bases, operated at CDS, Strasbourg, France. This publication makes use of data products from 2MASS, which is a joint project of the University of Massachusetts and the Infrared Processing and Analysis Center, California Institute of Technology also use of data product from the Wide-Field Infrared Survey Explorer, which is a joint project of the University of California, Los Angeles, and the Jet Propulsion Laboratory/California Institute of Technology, funded by the National Aereonautics and Space Administration. The authors wish to express their gratitude to Anahit Samsonyan for her help with Spitzer data analysis.